\documentclass[]{cDSS2e}

\begin{document}



\markboth{P. F. Bernath}{Molecular astronomy of cool stars and sub-stellar objects}


\title{{\textbf Molecular astronomy of cool stars and sub-stellar objects}}

\author{PETER F. BERNATH$^{\ast}$\thanks{$^\ast$ Email: pfb500@york.ac.uk \vspace{6pt}}\\\vspace{9pt} {Department of Chemistry, University of York, Heslington, York, YO10 5DD UK}\\\vspace{9pt}\received{v3.5 released January 2009} }

\maketitle
\bigskip
\begin{abstract}
The optical and infrared spectra of a wide variety of `cool' astronomical objects including the Sun, sunspots, K-, M- and S-type stars, carbon stars, brown dwarfs and extrasolar planets are reviewed. The review provides the necessary astronomical background for chemical physicists to understand and appreciate the unique molecular environments found in astronomy. The calculation of molecular opacities needed to simulate the observed spectral energy distributions is discussed.
\bigskip


\newpage

\centerline{\bfseries Contents}
\bigskip

{1.}    Introduction\\

{2.}    Astronomy Background\\

{3.}    The Sun and Sunspots\\

{4.}    K and M Stars\\

{5.}    Brown Dwarfs\\

{6.}   AGB, S and C Stars\\

{7.}   Extrasolar Planets\\

{8.}   Molecular Opacities\\

{9.}   Conclusions\\

{10.}   References\\
\end{abstract}

\section{Introduction}

Molecules are associated with the `cold' Universe. Historically astronomers first studied the spectra of hot, bright stars full of atomic emission and absorption lines. Except for the nearby Sun, cooler stars with molecular lines are fainter and more difficult to observe. As the surface temperature drops, the star reddens and the peak of the emission curve shifts to the infrared. (The Wien displacement law for a blackbody, $\lambda T =2898 \; \mu$m K, gives a maximum at 1 $\mu$m for 2898 K. \cite{book}) The total radiated power also varies approximately as the fourth power of temperature (from the Stefan--Boltzmann law \cite{book}), in so far as stellar emission is approximated by a blackbody. The spectroscopic study of `cool' sources with surface temperatures in the range of about $500-4000$ K is therefore best carried out with large modern 10 m class telescopes to gather as many photons as possible in the near infrared and infrared spectral domains.

Molecules are in fact ubiquitous in the Universe and are found in diffuse clouds, dark clouds, giant molecular clouds, planetary nebulae, circumstellar envelopes, stellar atmospheres, brown dwarf atmospheres, planetary atmospheres, comets, supernovae and galaxies. Molecules in dark clouds, circumstellar envelopes, and star-forming regions are usually studied by recording pure rotational spectra using radio telescopes because these long wavelengths are scattered less than infrared and optical wavelengths in these dust-obscured sources. Temperatures are very low, typically less than 50 K. There are already a number of good reviews on interstellar clouds and related objects from an astrochemical perspective \cite{herbst, vanD, lucy, vanD2} and they will not be covered here. Instead this review will focus on visible and infrared observations of molecules in cool stars, brown dwarfs and extrasolar planets using a mixture of spectroscopic and astronomical citations. The distinction between molecules found in stellar atmospheres and those in molecular clouds is not always clear because many stars are enshrouded by clouds of molecules and dust, which can then appear in the stellar spectra. For example, Mira variable stars have prominent hot water overtone absorption bands that originate from at least two pulsing circumstellar shells of water vapour \cite{hinkle}.

Modelling of the chemical composition of stellar atmospheres is also different from that of dark clouds. The very low temperatures of dark clouds mean that only reactions with very small energy barriers are feasible (e.g. ion-molecule reactions) and the source is not in chemical equilibrium \cite{herbst2}. The composition of a dark cloud is therefore calculated by assuming an initial composition and solving the coupled differential rate equations as is typical in chemical kinetics. In contrast, the higher temperatures and pressures in stellar atmospheres mean that the system has largely attained chemical equilibrium. The composition of a shell in a stellar atmosphere can therefore be predicted given an elemental composition, temperature ($T$), pressure and Gibbs free energy of formation ($\Delta G_f(T)$) of all possible compounds as a function of temperature \cite{tsuji}. In this case, the coupled algebraic equations that describe the equilibria need to be solved simultaneously to determine the composition, as is customary in chemical thermodynamics.

\section{Astronomy Background}

Big Bang nucleosynthesis created only the elements H, He and a trace of Li some 13.7 billion years ago \cite{schramm, wmap}; all other elements are formed in stars \cite{trimble, wallerstein}. After the Big Bang, the early Universe expanded rapidly and cooled. After a few hundred million years, gravitational attraction produced the first stars, the `Population III' class of objects \cite{bromm}. Little is known about this first generation of stars because they have not been detected yet with certainty; a few molecules such as LiH are predicted to form in the early Universe \cite{lepp}. Models predict that population III stars were massive (more than one hundred solar masses) and their intense UV emission ionised the surrounding interstellar medium. Massive stars burn the H and He fuel rapidly to form the elements up to Fe in the periodic table by nuclear fusion reactions. The Fe nucleus is the most stable in terms of binding energy per nucleon so fusion reactions are unable to make any heavier elements. After a few million years of life Population III stars began to collapse as the nuclear fuel was exhausted, and they exploded as supernovae. Nuclear reactions during supernova explosions release a rapid burst of neutrons (\textit{r}-process) that are absorbed by nuclei and make elements heavier than iron. Some heavy elements are also produced by nuclear reactions in stellar cores from the slow neutrons (\textit{s}-process) released in fusion reactions. Supernova explosions and stellar winds therefore return matter to the interstellar medium enriched in `metals'. In astronomy all elements other than H and He are termed metals.

The Population II class of stars were created by the gravitational collapse of matter in the interstellar clouds that formed from the metal-enriched ejecta of Population III stars. Population II stars are still depleted in metals compared to younger stars like our Sun that are part of Population I. The elemental abundance of a star is a crucial property and our own Sun is taken as a standard reference. Astronomers have defined the metallicity of a star as the abundance ratio of iron to hydrogen as compared to the same ratio in the Sun, i.e. the metallicity [Fe/H] is defined as

\begin{equation}
  [\mathrm{Fe/H}] = \log_{10} \left(\frac {(N_{\mathrm{Fe}}/N_{\mathrm{H}})_{\mathrm{star}}} {(N_{\mathrm{Fe}}/N_{\mathrm{H}})_{\mathrm{Sun}}}\right) = \log_{10}\left(\frac {N_{\mathrm{Fe}}}{N_{\mathrm{H}}}\right)_{\mathrm{star}} - \log_{10} \left(\frac {N_{\mathrm{Fe}}}{N_{\mathrm{H}}}\right)_{\mathrm{Sun}}
\end{equation}
so a metallicity of 0 corresponds to the same relative Fe abundance as the Sun. Fe is chosen for measurement convenience, and metallicities of $-1$ and $-2$ correspond to relative Fe depletions by factors of 10 and 100, respectively, compared to the Sun. Population II stars have metallicities more negative than $-1$ and Population III stars should have essentially no metals and metallicites less than $-6$, i.e. Fe depleted by more than a factor of a million, are expected. A few low mass stars with metallicites as small as $-5.3$ have been detected, but their evolutionary histories are not known with certainty \cite{christlieb}.

Normal stars are classified primarily by colour \cite{allen}. This so called `main sequence' contains the usual stellar types (O B A F G K M) classified by temperature from about 60\,000 K (blue, early O-type stars) to 2400 K (red, late M-type stars).  Each letter is divided into 10 subdivisions ($0-9$) so at 5800 K our Sun is a G2 dwarf. `Early' M's (or other stellar types) refer to low numbers and `late' M's to cooler higher numbers. The classification of stars is two dimensional, based both on surface temperature (e.g. G2 for 5800 K) and `size' (luminosity).  In practice there are three main categories for luminosity:  supergiant (I), giant (III) and dwarf (V), although five Roman numerals (I to V) are used. Luminosity (with $1\; L_\odot=3.845\times 10^{26}$ W equal to the luminosity of the Sun) is used for convenience instead of the more difficult to determine mass (in units of solar masses, $1\; M_\odot = 1.989\times 10^{30}$ kg). Dwarfs are the largest category and our Sun is thus classified as a G2 V star. Stellar interiors have temperatures of millions of degrees (e.g. 15 million K for our Sun), but they are not visible from the exterior because of the `opacity' of stellar atmospheres. Implicit in the stellar main sequence is the correlation between mass and surface temperature: massive stars burn their nuclear fuel more rapidly and have higher surface temperatures.

In practice stellar classification is based on the presence or absence of key atomic and molecular features in the spectrum \cite{kaler}, which is sometimes called the spectral energy distribution. A-type stars have strong hydrogen lines and the coolest stars have strong near infrared bands of electronic transitions of diatomic TiO and VO. Recently, objects cooler and smaller than M-type stars have been discovered and called brown dwarfs \cite{bd}. The classification system was therefore extended with two new spectral types, L and T; some early L's are stars, and late L's and T's are brown dwarfs. L-type objects have temperatures of about $2400-1300$ K and are characterised by the presence of near infrared electronic transitions of the metal hydrides FeH and CrH, and the fading of VO and TiO bands. T-type brown dwarfs have temperatures in the range of about $1300-700$ K and have strong absorption from overtone vibration-rotation bands of methane and water. Finally a new `Y' category of very cool brown dwarfs has been predicted with temperatures below about 600--700 K and characterised by strong ammonia overtone absorption bands \cite{ydwarf}.

Brown dwarfs are not stars because they are not massive enough to burn hydrogen in their cores, although nuclear fusion of other nuclei such as deuterium and lithium is still possible. Effective surface temperatures are still somewhat uncertain, but the new field of observation and modelling of sub-stellar objects is progressing rapidly.

The term `dwarf' is used in several different contexts in stellar astronomy. For example, the term subdwarf has little to do with luminosity but refers to a low abundance of heavy elements \cite{sd}. A subdwarf has a metallicity of at least $-1$ and an extreme subdwarf has a metallicity of $-2$. In fact the lack of metals in cool subdwarfs means that the opacity of the stellar atmosphere is decreased so that subdwarfs can be more luminous in the visible than dwarfs (i.e. subdwarfs are not necessarily `sub-luminous'). The terms dwarf, brown dwarf and subdwarf therefore refer to different kinds of objects.

Not all stars lie on the `main sequence' discussed above. Clearly very young stars---young stellar objects or YSOs---that have just been born in dark interstellar clouds do not lie on the main sequence. YSOs will evolve until they are main sequence stars and then spend most of their life on the main sequence \cite{vanD2}. Similarly when low-mass stars similar to our Sun convert most of their H nuclear fuel into He, they evolve off the main sequence and become red giants \cite{agb, sun}. Red giants burn He in their cores to make C, and their outer shells expand and cool. The red giant phase is followed by a complex AGB (asymptotic giant branch) phase where the core consists of mainly C and O with energy coming from the fusion of He and/or H in thin surrounding shells \cite{agb}. During the AGB phase strong stellar winds return much of the matter to the interstellar medium and the object evolves towards a planetary nebula. AGB stars often possess circumstellar shells of dust and molecules which envelop the star and render it nearly invisible in the optical region. In a planetary nebula the circumstellar shell has become detached from the hot central stellar core. A planetary nebula is a spectacular short-lived phase (a few thousand years) in which the surrounding cloud of matter glows in the visible from the intense UV excitation of the very hot central star. Ultimately the cloud of matter dissipates and the central star becomes a white dwarf.

High-mass stars with stellar masses greater than about 8 $M_\odot$ will explode as Type II supernovae, unless mass-loss is sufficiently rapid \cite{sn}. In essence, high-mass stars consume their nuclear fuel until a sufficiently large ($>1.3 \;M_\odot$, the Chandrasekhar limit) inert Fe/Ni core remains in which nuclear fusion is no longer possible. The heat production of the core is insufficient to support the outer layers and gravitational collapse ensues, rapidly releasing an enormous quantity of energy. The outer layers of the star disintegrate and ultimately the core ends up as a neutron star or a black hole.

The ratio of carbon to oxygen abundance, C/O, is a crucial parameter. Normal stars on the main sequence have C/O less than 1 and when molecules begin to form CO is always present because of its strong triple bond. The formation of CO ties up most of the carbon and an oxygen-rich chemistry results. If C/O is greater than 1 then the formation of CO ties up all of the oxygen and a very different carbon-rich chemistry is possible. Red giant and AGB stars produce carbon from helium so the composition of stars that are no longer on the main sequence becomes increasingly carbon rich. When the abundance of carbon and oxygen are equal then S-type stars (see below) result \cite{s-stars} and when carbon is more abundant than oxygen then carbon stars (C-type stars) form \cite{c-stars}. Carbon stars are therefore red giant or AGB stars with C/O $>$ 1.

Giant planets such as Jupiter ($1\; M_{\mathrm{J}}= 0.955\times10^{-3}\; M_\odot= 318\; M_\oplus$), which has a surface temperature less than 200 K, are believed to fall on the same main sequence as brown dwarfs. Hundreds of extrasolar planets have been discovered primarily by the very small periodic Doppler shifts they induce in the absorption lines of the spectrum of the parent star \cite{burrows, esp}. A few extrasolar planets have been detected by the small decrease in luminosity observed when they transit in front of the star. There is a selection effect that preferentially favours the discovery of `hot Jupiters' because massive planets with short periods are easier to find. Extrasolar planets are therefore closely related to brown dwarfs and will be included in the set of cool objects covered in this review.

\section{The Sun and Sunspots}

As our nearest and most important star, the Sun holds a special position in astronomy and its study is part of a separate sub-discipline. The Sun has an effective surface temperature of 5800 K, which dissociates all polyatomic molecules. Lines of neutral atoms and atomic ions dominate the solar spectrum as first observed by Fraunhofer in the visible region. Fraunhofer had no explanation for the dark absorption lines in the solar spectrum and labelled them with letters starting with A in the near infrared at 760 nm. Apart from the A- and B-lines (which are the forbidden A- and B-bands of terrestrial
O$_2$, i.e. the 0-0 and 1-0 vibrational bands of the $\mathrm{b}^1\Sigma_{\mathrm{g}}^+ - \mathrm{X}^3\Sigma_{\mathrm{g}}^-$ electronic transition), all of the Fraunhofer lines are due to atoms; for example, the C-line at 656 nm is the first H Balmer line (H$_\alpha$) and the D-line contains the two Na D-lines (589.0 and 589.6 nm).

A handful of diatomic molecules contribute weakly to the near infrared, visible and near UV solar spectra \cite{whl, whl2}: MgH $\mathrm{A}^2\Pi - \mathrm{X}^2\Sigma^+$ \cite{mgh1,mgh2}, C$_2$ Swan bands ($\mathrm{d}^3\Pi_{\mathrm{g}} - \mathrm{a}^3\Pi_{\mathrm{u}})$ \cite{swan}, CH $\mathrm{B}^2\Sigma^- - \mathrm{X}^2\Pi$ and $\mathrm{A}^2\Delta - \mathrm{X}^2\Pi$ \cite{ch}, CN $\mathrm{B}^2\Sigma^+ - \mathrm{X}^2\Sigma^+$ (Violet System) \cite{violet} and $\mathrm{A}^2\Pi - \mathrm{X}^2\Sigma^+$ (Red System) \cite{red}. Solar spectra in this region have been recorded with the 1-m Fourier transform spectrometer (FTS) at the McMath--Pierce Solar Telescope on Kitt Peak and published in a series of solar atlases by L. Wallace and co-workers \cite{whl, whl2}; there is also an earlier high resolution solar flux atlas \cite{flux}.

In the infrared, the solar spectra are more interesting as they contain prominent molecular bands. Excellent ground-based spectra are again available from the Kitt Peak FTS covering the 1850--9000 cm$^{-1}$ \cite{lw, lw2} and 460-- 630 cm$^{-1}$ regions \cite{wlb}. In addition, observations from Jungfraujoch in Switzerland (a higher, drier site than Kitt Peak) extend to longer wavelengths \cite{farmer} and cover 250--630 cm$^{-1}$. At short wavelengths the CN Red System and the C$_2$ Phillips System ($\mathrm{A}^1\Pi_{\mathrm{u}} - \mathrm{X}^1\Sigma^+_{\mathrm{g}}$) \cite{phillips} can be seen, but the spectra are dominated by the strong first overtone and fundamental vibration-rotation bands of CO near 2.5 $\mu$m and 5 $\mu$m \cite{co}. The Meinel bands (OH vibration-rotation bands) near 3 $\mu$m are also strong and the OH pure rotational lines are prominent at long wavelengths \cite{colin}. The vibration-rotation bands of CH can be seen at 3 $\mu$m \cite{ch, ch2}.

In the infrared, excellent high resolution solar spectra have also been recorded from low earth orbit by the ATMOS and ACE FTSs. The ATMOS instrument produced two solar atlases covering 625--4800 cm$^{-1}$ at 0.01 cm$^{-1}$ resolution \cite{atmos1, atmos2} and the ACE atlas covers 750--4400 cm$^{-1}$ at 0.02 cm$^{-1}$ resolution \cite{ace, hase} with a higher signal-noise ratio and no telluric interference from residual gases in the instrument, Figure 1 (http://www.ace.uwaterloo.ca/solaratlas.html). The satellite spectra also clearly show vibration-rotation \cite{nh} and pure rotational lines of NH. The combination of laboratory and solar spectra have led to a considerable improvement in the spectroscopic constants of OH \cite{colin}, NH \cite{nh}, and CH \cite{ch} in their electronic ground states.

\begin{figure}
\begin{center}
\rotatebox{270}{\resizebox{!}{14cm}
{\includegraphics{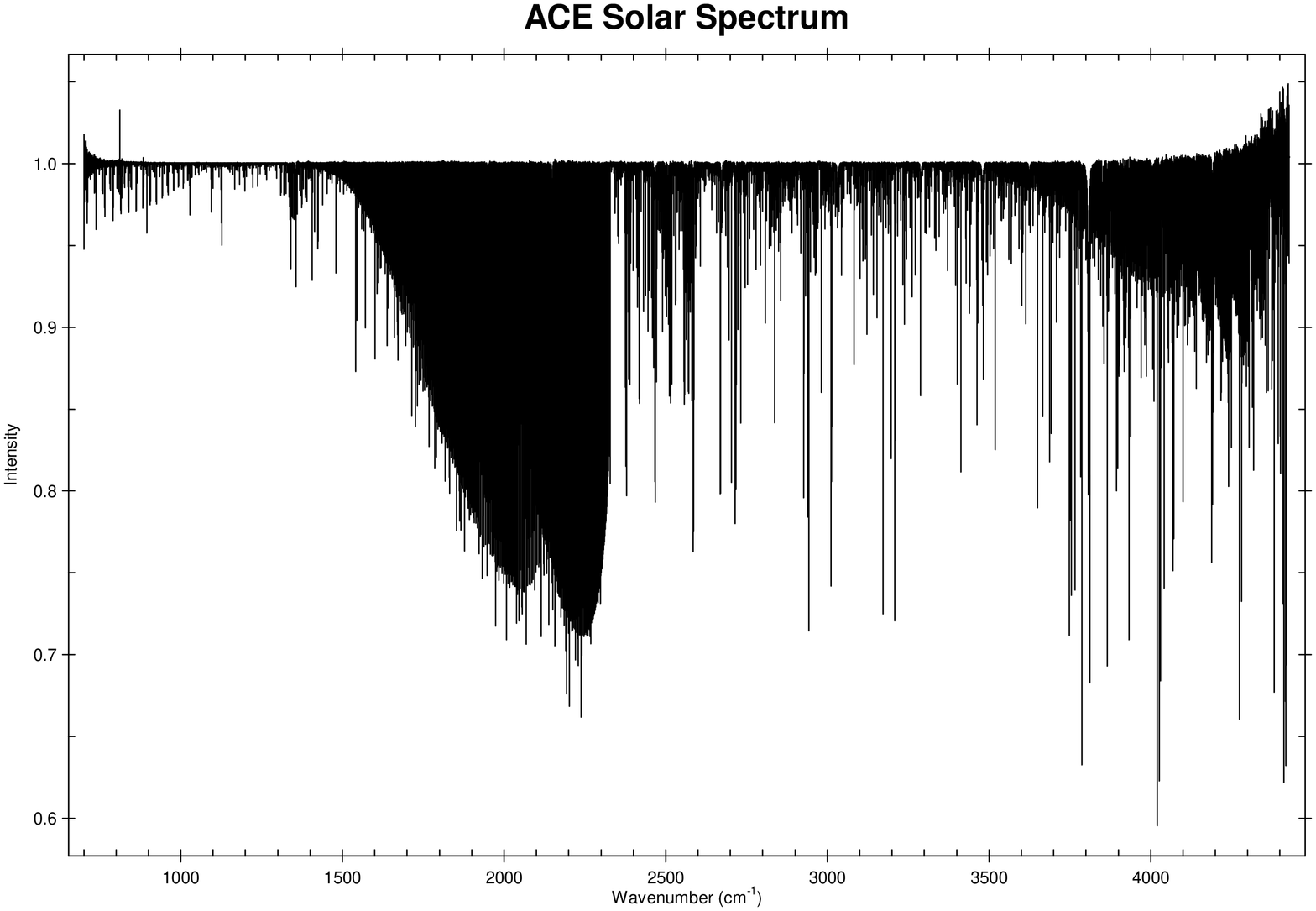}}}
\caption{The ACE solar spectrum with the intensity scale normalised to 1. Strong CO absorption is seen near 2000 and 4000 cm$^{-1}$ and OH pure rotational lines near 1000 cm$^{-1}$ are clear.}
\end{center}
\end{figure}

\begin{figure}
\begin{center}
{\resizebox{!}{10cm}
{\includegraphics{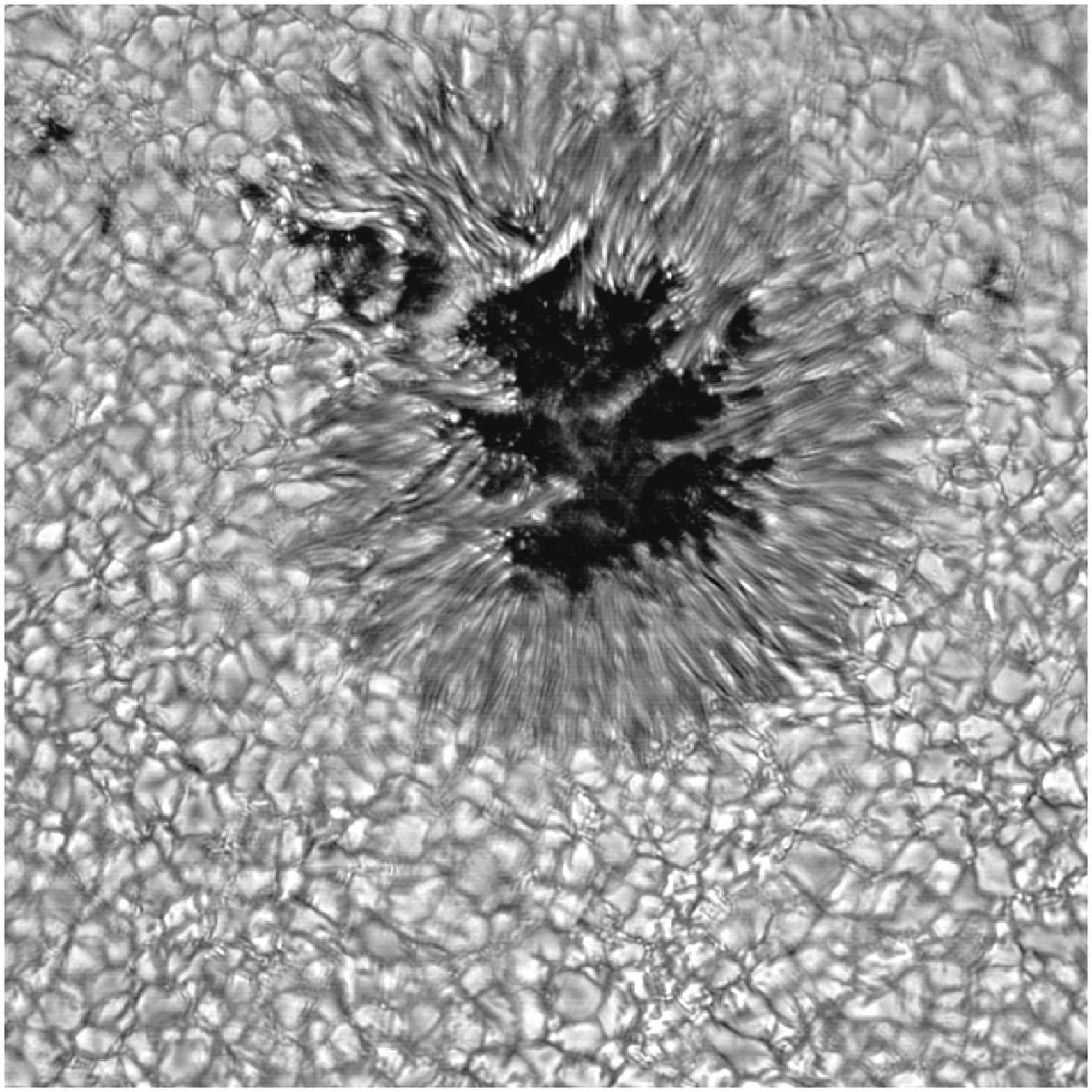}}}
\caption{An image of a sunspot as seen with the Vacuum Tower Telescope of the National Solar Observatory at Kitt Peak.}
\end{center}
\end{figure}

Sunspots are cooler than 5800 K---hence their apparent contrasting black colour. Large black sunspots (Figure 2) can be as cool as 3300 K, equivalent to an M4 star, but with a vertical magnetic field of more than 0.3 T. There are a series of Kitt Peak atlases for sunspots that cover 470--1233 cm$^{-1}$ \cite{wlb}, 1970--8640 cm$^{-1}$ \cite{wl}, 1925--3480 cm$^{-1}$ \cite{whl-2}, 4000--8640 cm$^{-1}$ \cite{whl3}, 8900--15\,050 cm$^{-1}$ \cite{wlbr}, and 15\,000--23\,000 cm$^{-1}$  \cite{whl4}. Paper copies are available from L. Wallace (wallace@noao.edu) and electronic copies from ftp://nsokp.nso.edu/pub/atlas/.

All of the molecules seen in the photosphere intensify in the spectra of sunspots, with the exception of C$_2$, which fades and is not obvious. The weak MgH $\mathrm{B^\prime}^2\Sigma^+ - \mathrm{X}^2\Sigma^+$ \cite{mgh3, mgh2} transition makes its appearance in the near infrared, along with the $\mathrm{A}^1\Pi - \mathrm{X}^1\Sigma^+$ transition of AlH \cite{alh} near 425 nm and the $\mathrm{B}^2\Sigma^+ - \mathrm{X}^2\Sigma^+$ and $\mathrm{A}^2\Pi - \mathrm{X}^2\Sigma^+$ transitions of CaH in the red \cite{cah}. The MgH lines can be used to determine $^{24}$Mg:$^{25}$Mg:$^{26}$Mg isotopic abundance ratios \cite{mgh3}. The lines of several electronic transitions of TiO \cite{tio3} become extremely dense in the visible with the $\gamma$-bands ($\mathrm{A}^3\Phi - \mathrm{X}^3\Delta$) \cite{tio1} and $\delta$-bands ($\mathrm{b}^1\Pi - \mathrm{a}^1\Delta$) \cite{tio2} in the red and near infrared. The Sun and sunspots are remarkable sources for spectroscopy at 3000--6000 K, and the two TiO papers by Ram et al. \cite{tio1, tio2} nicely demonstrate how laboratory and solar data (Figure 3) can be combined to give improved spectroscopic constants.

\begin{figure}
\begin{center}
\rotatebox{270}{\resizebox{!}{15cm}
{\includegraphics{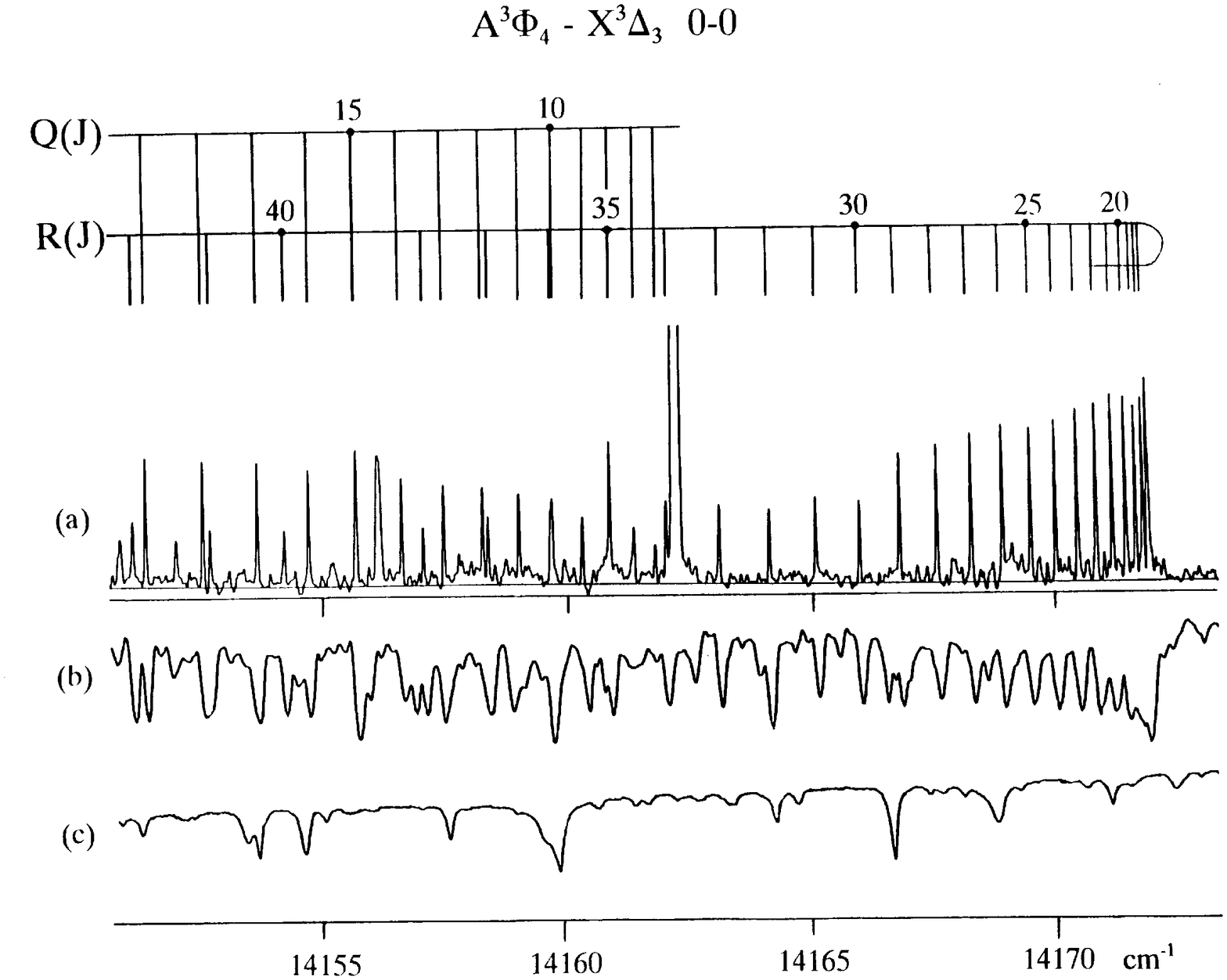}}}
\caption{The 0-0 $\delta$-band of TiO as (a) seen in the laboratory in emission as (b) seen in a sunspot in absorption and (c) absent in the solar photosphere \cite{tio2}. Reproduced by permission of the AAS.}
\end{center}
\end{figure}

As in M-type stellar spectra the FeH $\mathrm{F}^4\Delta - \mathrm{X}^4\Delta$ transition \cite{feh1, feh2} near 1 $\mu$m and the $\mathrm{E}^4\Pi - \mathrm{A}^4\Pi$ transition \cite{balfour} near 1.58 $\mu$m are present in sunspots, but with a large Zeeman effect. The FeH F-X 0-0 band near 990 nm is called the Wing--Ford band in the astronomical literature (see below). Interestingly the sunspot spectra were combined with laboratory observations to derive rotational \emph{g}-values for levels involved in some of the low-\emph{J} lines in the Wing--Ford band \cite{harrison}. These \emph{g}-values are useful in measuring magnetic fields in cool objects. For hotter objects such as the Sun and sunspots the magnetic field strength is obtained from magnetically-sensitive atomic lines (i.e. lines between levels with large \emph{g}-values) either by the observation of Zeeman splittings for large fields or from the polarization of the lines for weaker fields. In cooler stars and brown dwarfs there are no useful atomic lines, so it has been proposed to use the Zeeman effect on the lines in the Wing--Ford band instead.

The infrared spectra of sunspots display a great profusion of lines: in addition to the CO, OH, NH and CH molecules seen in the photosphere, SiO \cite{sio}, HF \cite{hf}, HCl \cite{hcl}, and H$_2$O \cite{water1, water2} appear. The fundamental and first overtone bands of SiO cover a substantial part of the 10 $\mu$m and 5 $\mu$m regions \cite{sio}, and H$_2$O has an amazing density of up to 50 lines/cm$^{-1}$ starting with the pure rotational region at the long wavelength limit and extending to nearly 7000 cm$^{-1}$ \cite{water3} in the near infrared.

An important advance in the spectroscopy of hot water occurred as a
result of the detection of water vapour in sunspots.  At 5800 K, the
Sun's photosphere is too hot for water to exist, but by 3900 K the concentration of OH and H$_2$O are equal.  A large number of unassigned lines were noticed in two Kitt Peak sunspot atlases
\cite{wl,wlb}. It was suspected that these lines were due to hot
water but the available laboratory data were inadequate to confirm
this. New laboratory infrared emission spectra of hot
water at 1800 K were therefore recorded. Comparison of the laboratory
emission spectra of H$_2$O and the sunspot absorption spectra
identified most of the unassigned sunspot lines as H$_2$O lines, but
only a small fraction of the new water lines could be assigned
quantum numbers. This work proved that there was `water on the
Sun' but the line density was so high that the spectrum seemed
`unassignable' by conventional techniques \cite{water1}.

We began a collaboration with the theoreticians Polyansky and
Tennyson to apply more sophisticated approaches to the problem
\cite{water2}. Through variational calculations of the energy
levels using a high quality {\it ab initio\/} potential energy surface, Polyansky et al.
\cite{water2} were able to assign most of the strong lines.  The
assignment method relies on the smooth variation of errors and
confirmation of tentative assignments by combination differences.

More recently, we have recorded a laboratory spectrum of water
vapour at 3000 K by using an oxy-acetylene torch as a source. The
water emission in the 500--13\,000 cm$^{-1}$ spectral region was
recorded with a high-resolution FTS \cite{water4}. Work
on this 3000 K spectrum of H$_2$O has just finished with the publication of the last paper in the series covering the near infrared part \cite{water5}. This spectrum matches
closely the conditions encountered in the photospheres of late M
dwarfs and sunspots. A paper on `monodromy', that is, the
change in the bending energy level pattern observed when the
highly-excited water molecule starts to sample linear geometries,
has also been published \cite{water6}. In this paper, we identify
vibration-rotation energy levels for the highly excited bending
states 8$\nu_2$ and 9$\nu_2$ above the barrier to linearity for the
first time. In our analysis we included sunspot spectral data.

An important application of hot water work is the calculation of
molecular water opacities for various cool objects.  Reliable
potential and dipole surfaces are needed to compute the millions of
transitions required to reproduce low-resolution infrared
spectra of M-type stars and brown dwarfs. The latest water line list
of Barber et al. \cite{bt2} contains more than 500 million lines and is
recommended for the simulation of the spectral energy distribution
of cool stars and brown dwarfs.

Recently the solar abundances of C, N, O, and Ne have been revised downwards by nearly a factor of two based on the intensity of atomic and molecular lines through the use of sophisticated 3-dimensional radiative hydrodynamic model atmospheres \cite{asplund1, asplund2, asplund3, asplund4}. The metallicity of the Sun was decreased from the previously accepted value of about 0.0245 down to 0.0165 \cite{asplund4} (for the Sun, the metallicity is defined as the mass fraction of the elements that are not H and He).

The granulation (outside of the sunspot in Figure 2) seen in solar images is caused by the upward motion of matter in the hotter centre of a gas cell (`granule') and downward motion at the cooler edges. The complex fluid motion in the solar atmosphere leads to unusual asymmetric line shapes and to line shifts. The line width is larger than expected based on the Doppler formula \cite{book} because of mass motion, traditionally attributed to `micro' and `macro' turbulence \cite{dravins}. In fact modern radiation-hydrodynamic models are capable of reproducing the observed line shapes \cite{asplund5}, but it is not clear yet whether the new abundances are correct \cite{ayres}. The lower abundances seem to be inconsistent with models of the Sun \cite{asplund3} used in the relatively new field of helioseismology \cite{helio} that observes periodic Doppler shifts of solar lines. These Doppler shifts are due to the propagation of sound waves in the Sun and have a typical period of 5 minutes.

\section{K and M Stars}

The Sun and particularly sunspots cover much of the temperature range spanned by K (4000--5200 K) and M (2400--3700 K) stars. Perhaps the best published spectrum is that of Arcturus ($\alpha$-Bo\"{o}tis), one of the brightest stars in the northern sky \cite{arc1, arc2, arc3}. Arcturus is a K1.5 IIIpe red giant star with a surface temperature of 4320 K (pe stands for `peculiar emission' because of the presence of emission lines in addition to the usual photospheric absorption features). The infrared and near infrared atlas ($0.9-5.3\;\mu$m) of Arcturus was recorded with the FTS that was associated with the 4-m Mayall Telescope \cite{arc1} at Kitt Peak, the visible atlas (373--930 nm) with the 0.9-m coud\'{e} feed telescope and the coud\'{e} spectrometer at Kitt Peak \cite{arc2} and the UV atlas (115--380 nm) recorded mainly with the STIS instrument on the Hubble Space Telescope \cite{arc3}. As expected, based on the solar atlases, the same molecular features are present with the exception of C$_2$, and the Lyman and Werner bands of H$_2$ \cite{h2} appear in emission in the UV \cite{arc3}.

M stars are particularly important because they are the most numerous type of star in our Milky Way galaxy, with 70\% by number and 40\% by mass \cite{bean}. Because they are so faint, it is only relatively recently that a full range of M dwarf spectra have become available \cite{mdwarf1, mdwarf2}. The visible and near infrared spectra of M-type stars are dominated by TiO absorption bands (Figure 4) as discussed above for sunspots. In Figure 4, the terms `active' and `inactive' refer to magnetic activity which induces atomic emission lines such as Balmer H$_\alpha$ at 656 nm, and is associated with stellar rotation \cite{reiners1}. Notice that astronomers have retained the old atomic notation with Na I referring to the neutral Na atom and Ca II to Ca$^+$.

The TiO 0--0 $\mathrm{A}^3\Phi - \mathrm{X}^3\Delta$ band head appears near 705.4 nm in the M1 spectra (3700 K) and has nearly become saturated by M6 (2750 K) with weaker TiO features such as the 0--1 band at 758.9 nm now present \cite{leggett}. By M6 the $\mathrm{B}^4\Pi - \mathrm{X}^4\Sigma^-$ 1--0 and 0--0 bands \cite{vo1,vo2} of VO are present near 733 nm and 785 nm (Figure 4) and strengthen for late M objects (Figure 5) \cite{bd}. As in sunspots MgH ($\mathrm{B^\prime}^2\Sigma^+ - \mathrm{X}^2\Sigma^+$), CaH ($\mathrm{B}^2\Sigma^+ - \mathrm{X}^2\Sigma^+$ and $\mathrm{A}^2\Pi - \mathrm{X}^2\Sigma^+$ at 635 and 690 nm, Figures 4 and 5 ) and FeH ($\mathrm{F}^4\Delta - \mathrm{X}^4\Delta$) are all prominent. Perhaps the most interesting new feature (Figure 4) in M spectra is ascribed to the CaOH $\mathrm{\widetilde{A}}^2\Pi - \mathrm{\widetilde{X}}^2\Sigma^+$ transition at 625 nm \cite{caoh1, caoh2}. In the infrared the overtone spectra of water appear more strongly than in sunspots for the late M's (Figure 6).

\begin{figure} 
\begin{center}
{\resizebox{!}{10cm}
{\includegraphics{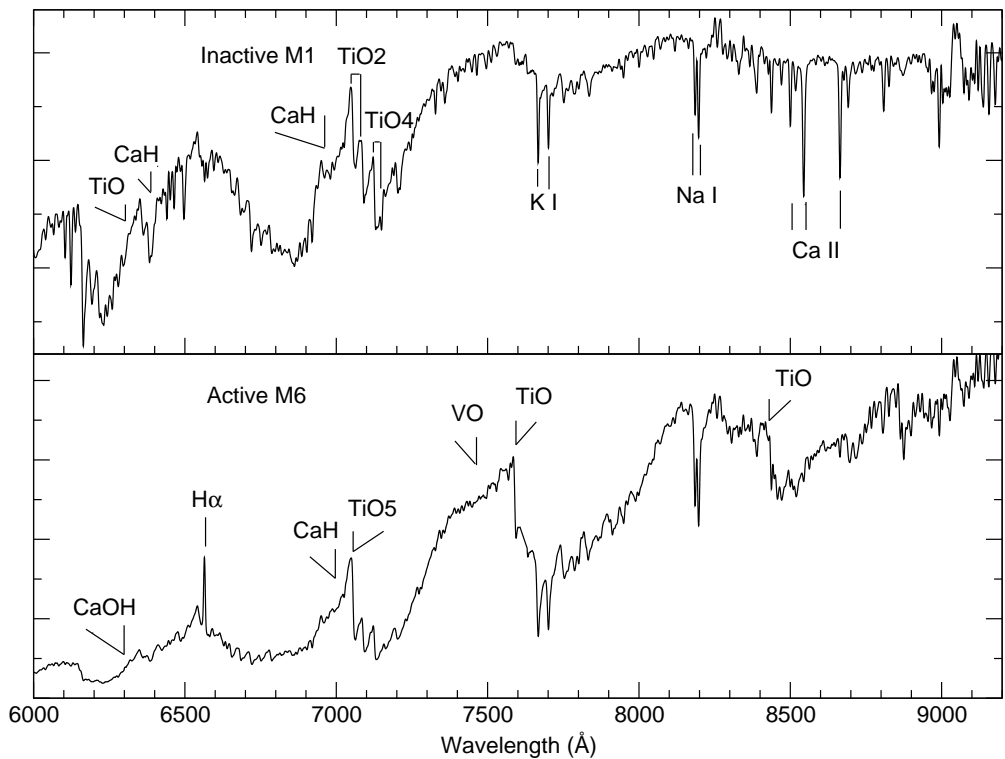}}}
\caption{Red and near infrared M dwarf spectra \cite{mdwarf2}. Reproduced by permission of the AAS.}
\end{center}
\end{figure}

In fact it was in late M dwarfs that Wing and Ford \cite{wing1} were the first to detect a
mysterious band near 991 nm at low resolution.  The
Wing--Ford band was later detected in S-type stars and in sunspots at
higher spectral resolution \cite{wing2}.  Nordh et al. \cite{wing3} identified
the Wing--Ford band as the 0-0 band of a FeH electronic transition
by comparison with an unassigned laboratory spectrum that had a band
head at 989.6 nm. The 1-0, 2-0, 2-1 and 0-1 bands of this transition
can all be identified in sunspot spectra. The spectra of the
Wing--Ford band and the other bands are very irregular
in appearance because of many overlapping branches with perturbed
lines.  Quantum chemistry predicts a large number of low-lying
electronic states and their mutual interaction causes perturbations
that appear as irregular patterns of lines. These `many-line
spectra' with no obvious patterns at first sight are characteristic
of the electronic spectra of transition metal hydrides.  After
heroic efforts, the near infrared transition of FeH was shown to be
a F$^4\Delta$--X$^4\Delta$ transition and seven bands with
v$^\prime$, v$^{\prime\prime}$ $\leq$ 2 were assigned by Phillips et al.
\cite{feh1} These bands were recorded in emission using the FTS at the National Solar Observatory at Kitt Peak using a carbon tube furnace source at about 2300$^\circ$C.  More recently the line intensities for the F$^4\Delta$--X$^4\Delta$ transition were obtained through \emph{ab initio} calculation \cite{feh2}.

\begin{figure}
\begin{center}
{\resizebox{!}{13cm}
{\includegraphics{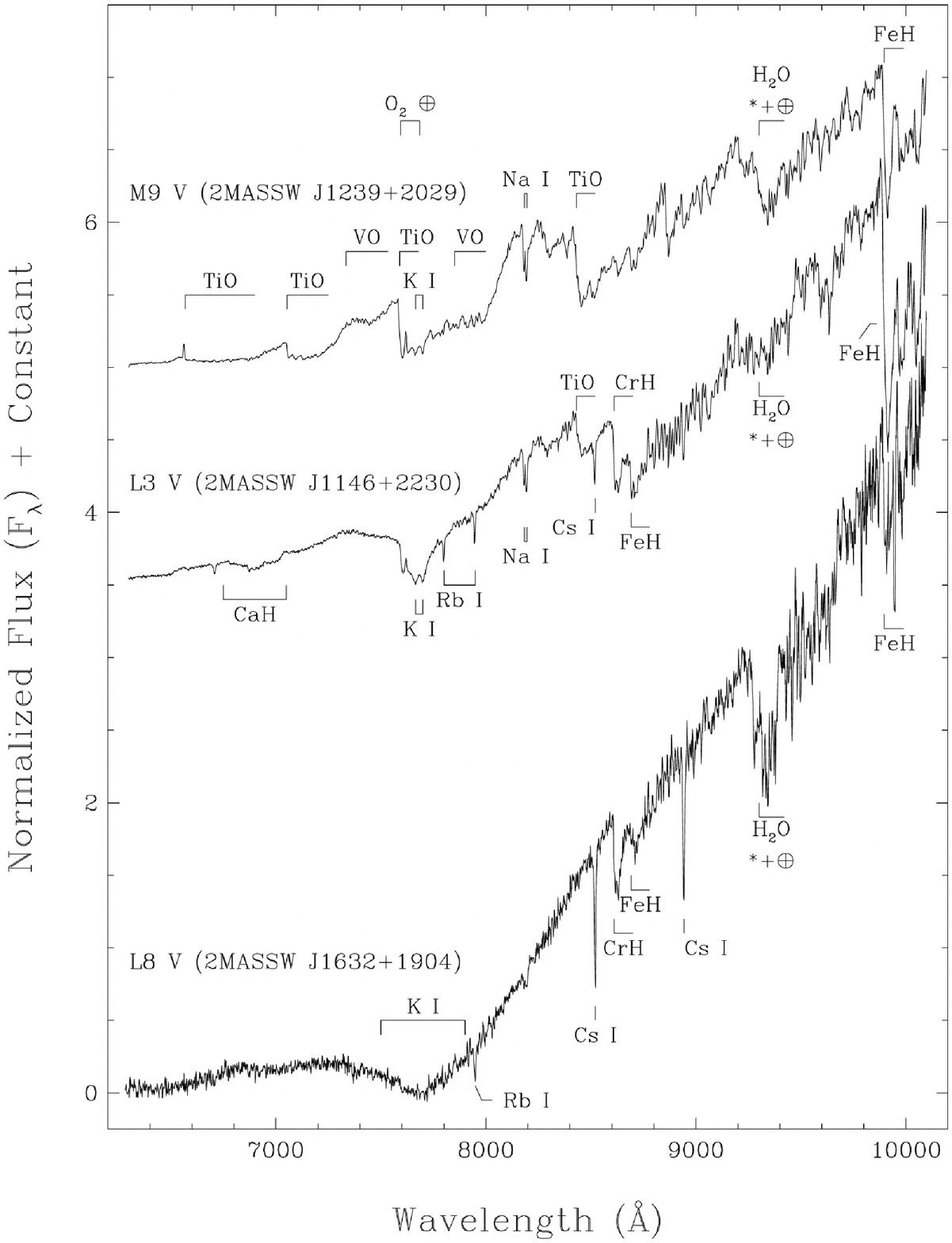}}}
\caption{Red and near infrared M9, L3 and L8 dwarf spectra \cite{ldef}. Reproduced by permission of the AAS.}
\end{center}
\end{figure}

The bands of CaH are particularly useful in identifying M subdwarfs that have reduced metal abundances compared to our Sun \cite{gizis}.  TiO and VO which characterise M-stars contain two heavy elements (`metals' in the astronomical sense), while CaH has only a single heavy element.  If the metal abundances are low, then the CaH bands strengthen relative to TiO, and the ratio of CaH to TiO band intensities can be used to identify subdwarfs \cite{gizis}.

\begin{figure}
\begin{center}
\rotatebox{90}{\resizebox{!}{14cm}
{\includegraphics{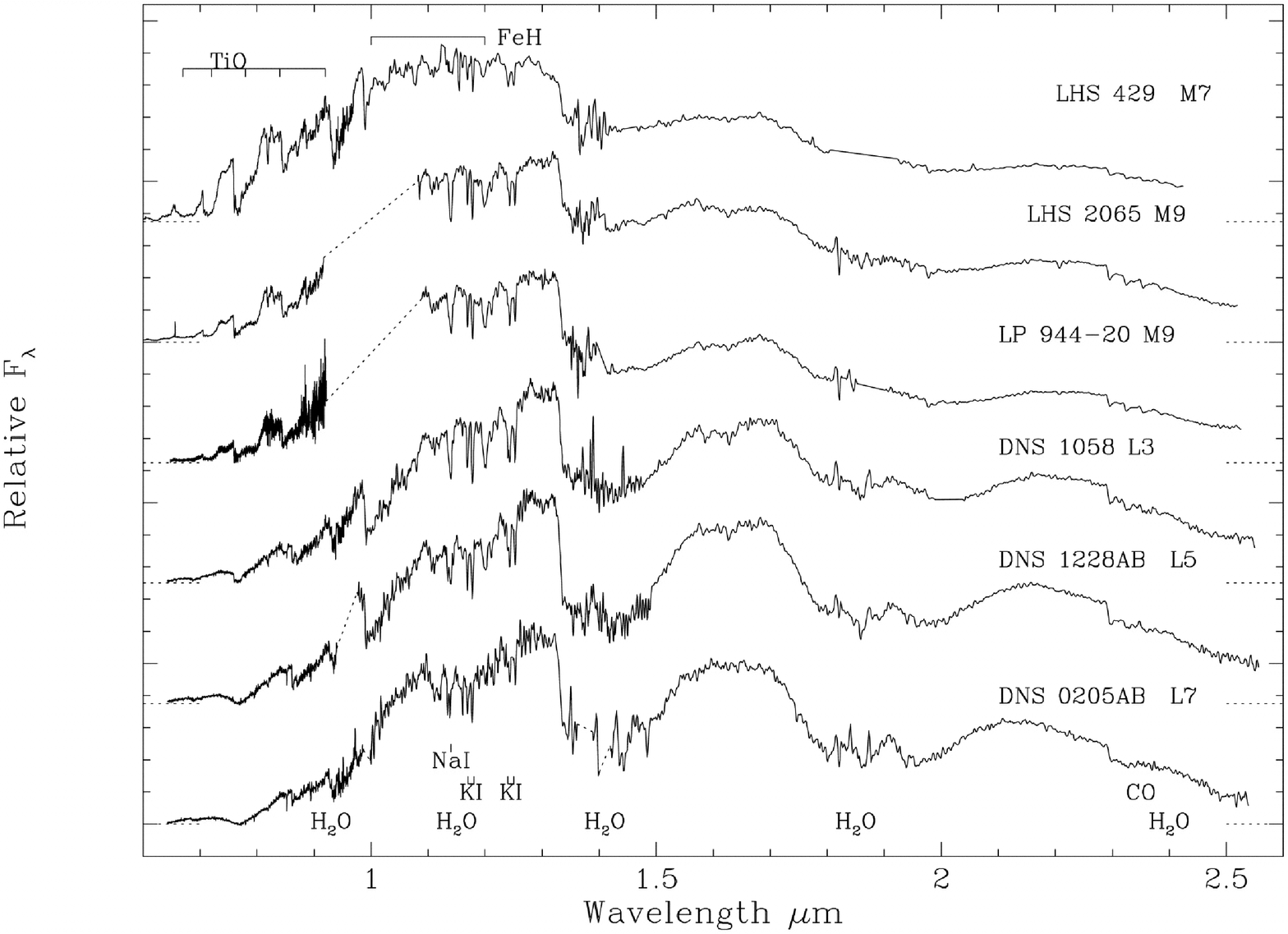}}}
\caption{Near infrared M and L dwarf spectra \cite{leggett-ldwarf}. Reproduced by permission of the AAS.}
\end{center}
\end{figure}

\begin{figure}
\begin{center}
{\resizebox{!}{13cm}
{\includegraphics{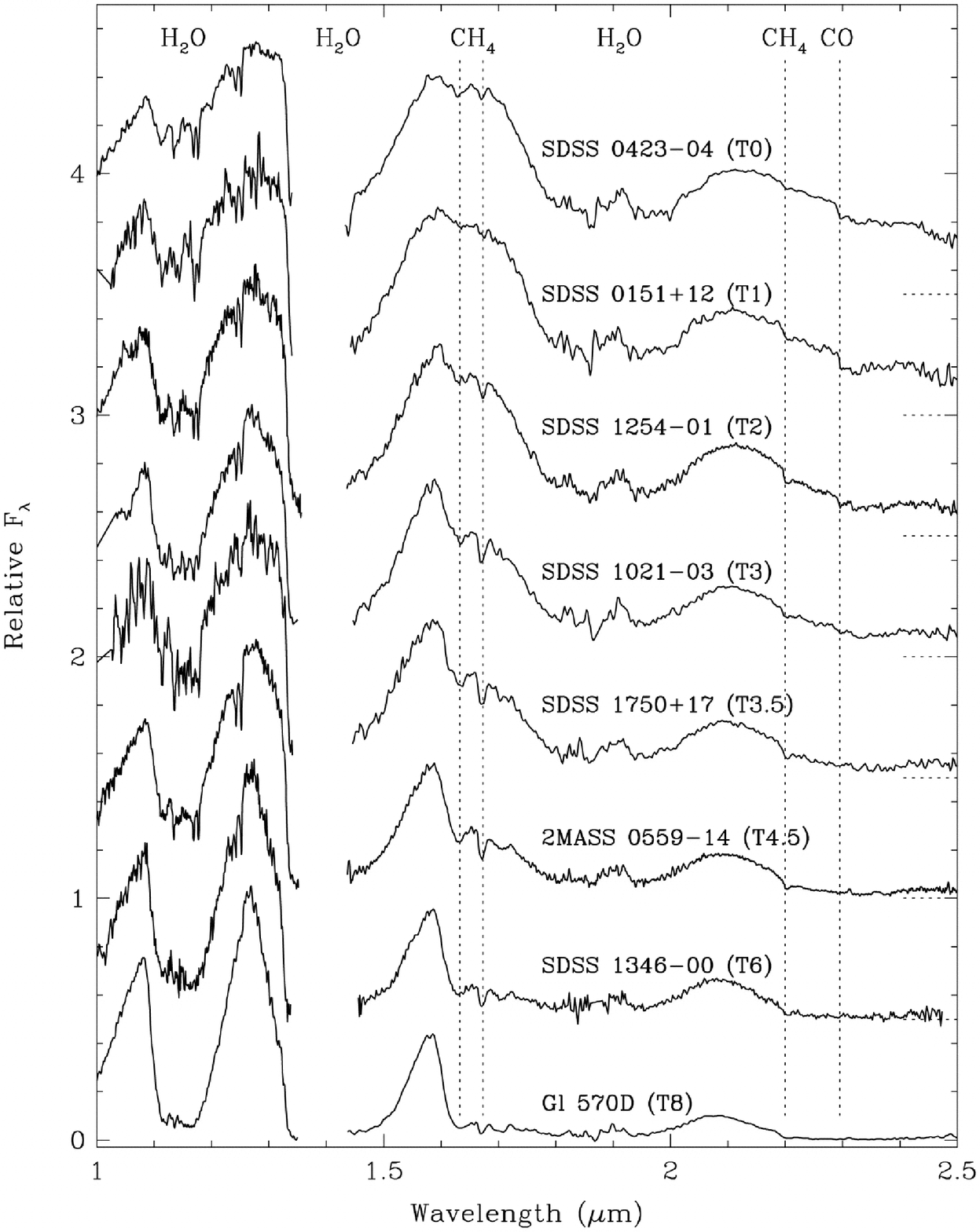}}}
\caption{Near infrared T dwarf spectra \cite{geballe}. Reproduced by permission of the AAS.}
\end{center}
\end{figure}

\section{Brown Dwarfs}

The discovery of the first brown dwarf Gl 229B by Oppenheimer et al. \cite{bd1} in 1995 was a momentous event and now hundreds of substellar objects are known (http://DwarfArchives.org/). Brown dwarfs are actually defined by mass because objects with less mass than about $0.075\; M_\odot=79\; M_{\mathrm{J}}$ ($1\; M_\odot=1047\; M_{\mathrm{J}}$, mass of Jupiter) are unable to fuse hydrogen in their cores but can still burn deuterium and lithium \cite{burrows2}. There is a similar deuterium burning limit at about $13\; M_{\mathrm{J}}$ and the lithium limit is about $63\; M_{\mathrm{J}}$. Cool objects in the L and T classes, however, are classified primarily by effective surface temperature, which controls the appearance of the spectra, and not by their difficult-to-determine mass. While there is certainly a correlation between mass and surface temperature, the relationship is not simple because of the different possible ages of the objects. Low mass objects burn hydrogen, deuterium, lithium, etc. slowly and can therefore have lifetimes of billions of years, and indeed some lifetimes are calculated to be comparable to the age of the Universe. It is therefore possible for a small young brown dwarf to have the same surface temperature (and spectral class) as a larger old low-mass star that has been slowly cooling.

The L class of objects is defined by the weakening of the VO and TiO metal oxide bands and the strengthening of the FeH and CrH metal hydride bands (Figure 5) \cite{bd,ldef}. The CrH A$^6\Sigma^+$--X$^6\Sigma^+$ transition with a 0--0 band head at 861 nm first appears in late M's (Figure 5) and is probably not present in sunspots \cite{crh1, crh2}. For later L's, the absorption bands of CrH and FeH fade and are replaced by infrared absorption of H$_2$O and CH$_4$ overtone and fundamental vibration-rotation bands (Figures 6 and 7) \cite{bd,tdef}.

The determination of surface temperatures of brown dwarfs is more complicated than for stars. In part the problem is due to their faintness and to difficulties observing in the infrared with ground-based telescopes. As Figure 7 illustrates, the spectral energy distributions no longer resemble blackbody curves---molecular absorption has chopped the spectral energy distributions into pieces. Surface temperatures therefore need to be determined using simulations and these models are still relatively primitive with incomplete molecular opacities. The transition from M to L objects is particularly difficult to model because of the condensation of gases to form dust; the L to T transition is easier to handle because the dust has settled out and has disappeared. In L-type objects particle scattering needs to be included in the simulation of spectral energy distributions in order to match the observed extinction. Existing effective surface temperatures are $\thicksim$2500 K for L0 and $\thicksim$1500 K for L9 while T0 start at $\thicksim$1500 K and descends to $\thicksim$800 K by T8 \cite{bd}.

Brown dwarf masses are also difficult to measure. Evolutionary models predict that while L's could be stars or brown dwarfs, T's are always brown dwarfs. One possible independent way to determine mass is from the surface gravity; astronomers are often able to determine the temperature, metallicity (composition) and gravity from high resolution stellar spectra. Gravity affects the observed spectra by changing the thickness and structure of the photosphere, which changes the appearance, for example, of the wings of lines. The first high resolution spectra are starting to appear \cite{reiners2, tennyson}, but one problem is that all known brown dwarfs are rotating rapidly with velocities in excess of 20 km/s, which corresponds to a spectral line broadening due to the Doppler effect of about 1 cm$^{-1}$ near 1 $\mu$m \cite{rotors}.

The importance of high spectral resolution is illustrated for TiH.  Opacity calculations predict that TiH should be particularly abundant for late M-type subdwarfs \cite{tih1}.	In 2003 Burgasser et al. \cite{burgasser1} and Lepine et al. \cite{lepine} classified two objects (2MASS 0532+8246 and LSR 1610-0040, respectively) as metal-deficient L-type subdwarfs.  In the Burgasser et al. paper \cite{burgasser1} there is an `unassigned hydride' band at 960 nm and another unassigned band at 940 nm.  Both of these features are overlapped by a water overtone band in the Earth's atmosphere and hot water absorption in the objects themselves. In a later paper on a third L-type subdwarf (2MASS 1626+3925) Burgasser \cite{burgasser2} suggests that the feature at 940 nm might be due to the A$^4\Phi$--X$^4\Phi$ transition of TiH \cite{tih2}.  Reiners and Basri \cite{reiners3} then recorded higher resolution spectra (resolving power, $\lambda/\Delta\lambda$, of 31000) of two of the three L-subdwarfs.  They conclude that LSR 1610-0040 is not an L-type subdwarf but a peculiar slightly metal-deficient M-type dwarf with strong CaH bands and relatively weak CrH and FeH features.  The feature near 960 nm is due to a group of Ti lines and there is no sign of the TiH lines near 940 nm in the high resolution spectrum of the L subdwarf 2MASS 0532+8246.

\begin{figure}
\begin{center}
{\resizebox{!}{13cm}
{\includegraphics{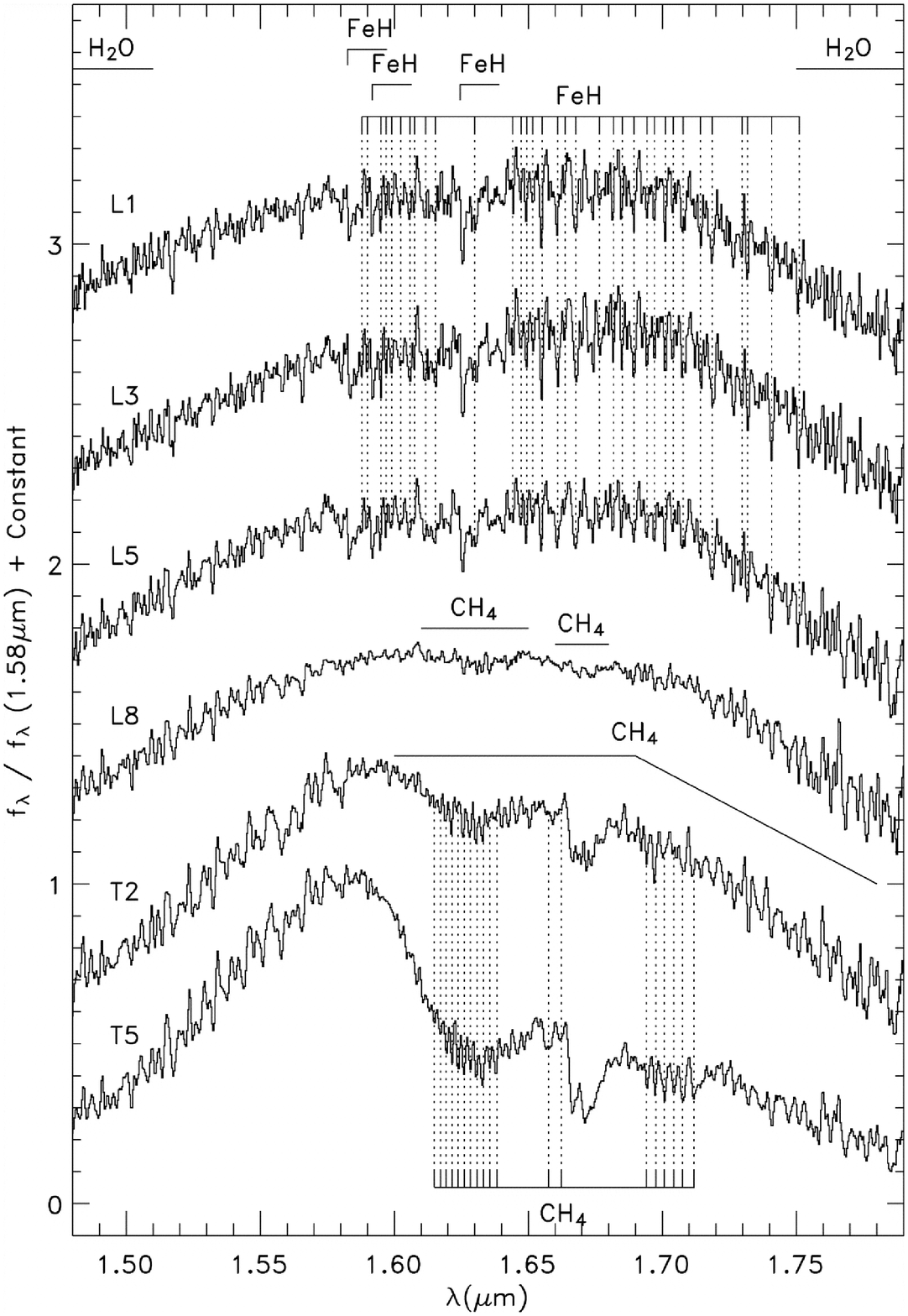}}}
\caption{Spectra of L dwarfs showing the FeH E-A transition and of T dwarfs with the methane C--H first overtone band \cite{cushing}. Reproduced by permission of the AAS.}
\end{center}
\end{figure}

\begin{figure}
\begin{center}
{\resizebox{!}{13cm}
{\includegraphics{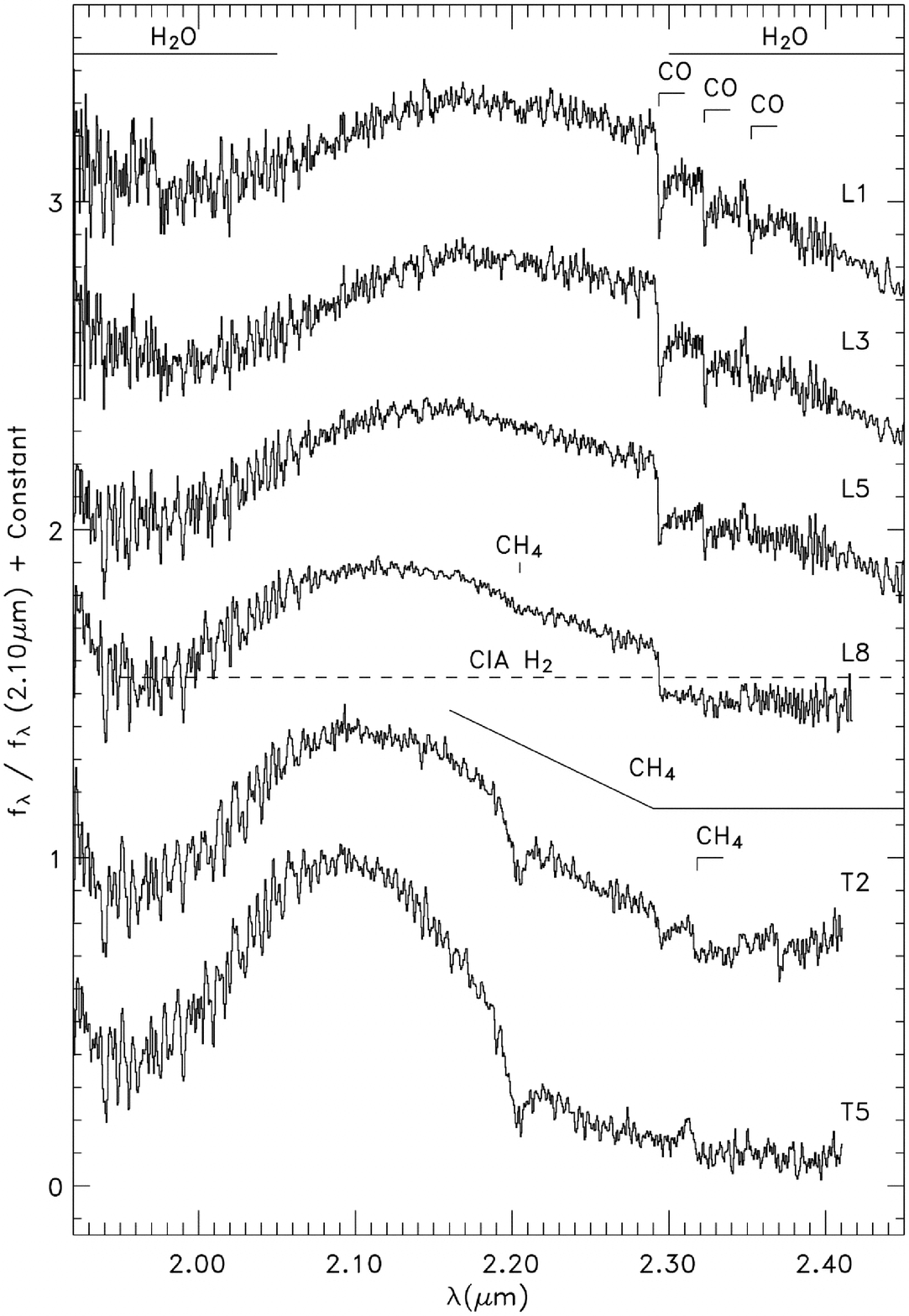}}}
\caption{Near infrared L and T dwarf spectra showing CO first overtone bands \cite{cushing}. Reproduced by permission of the AAS.}
\end{center}
\end{figure}

\begin{figure}
\begin{center}
{\resizebox{!}{13cm}
{\includegraphics{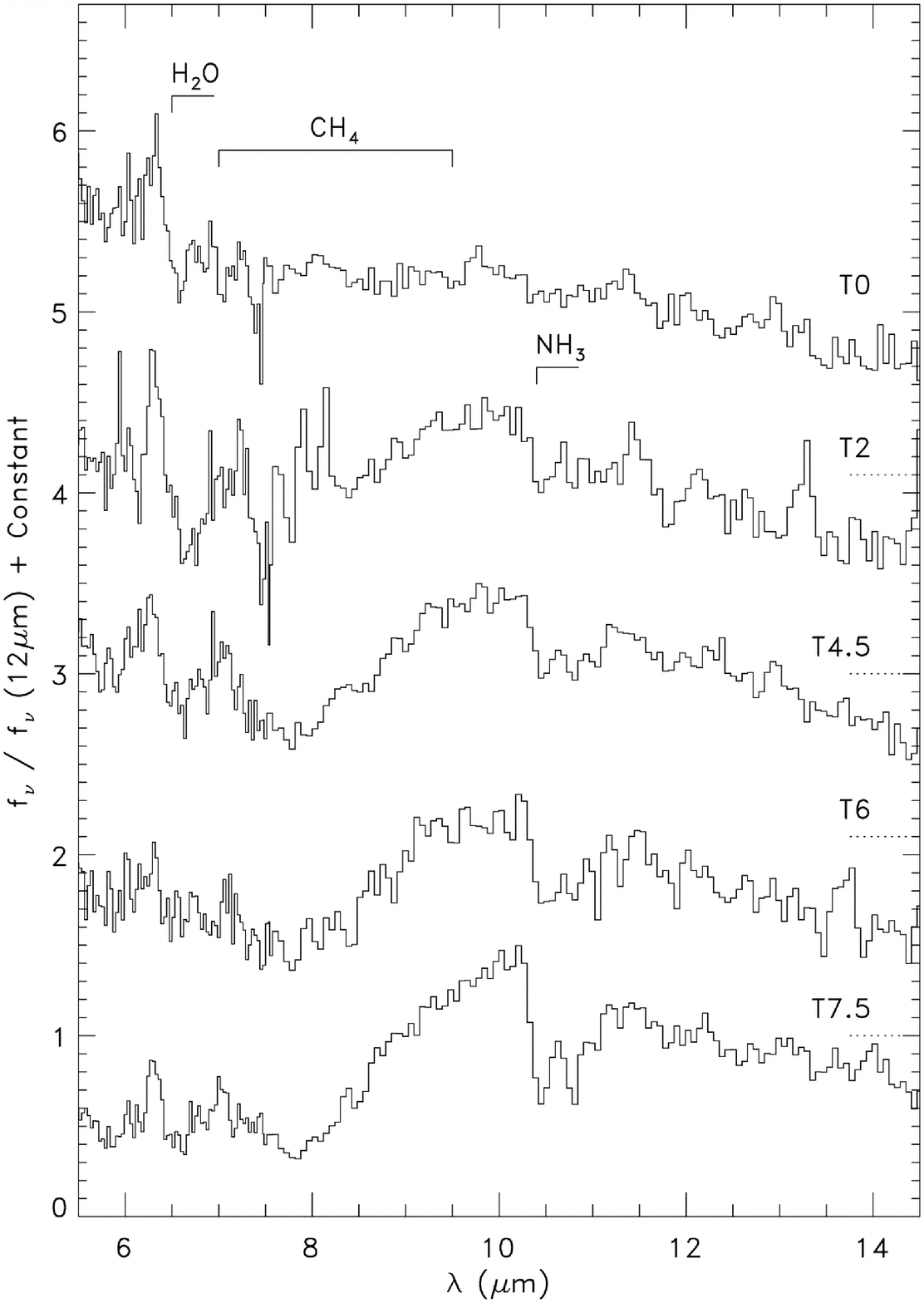}}}
\caption{Infrared spectra of L and T dwarfs with H$_2$O $\nu_2$ band near 6 $\mu$m, CH$_4$ $\nu_4$ band near 8 $\mu$m and NH$_3$ $\nu_2$ band near 11 $\mu$m \cite{cushing2}. Reproduced by permission of the AAS.}
\end{center}
\end{figure}

For T-dwarfs the main contributors to the opacity are water, methane and ammonia. Cushing et al. \cite{cushing} recorded a nice sequence of moderate resolution spectra of M, L and T dwarfs in the 0.6 to 4.1 $\mu$m region with ground-based telescopes. These spectra show beautiful methane features at 1.65 $\mu$m (2$\nu_3$, first overtone of the C--H stretch, Figure 8) that begin to saturate at 3.4 $\mu$m ($\nu_3$, C--H stretch fundamental) for a T5 dwarf \cite{nassar}. Near 2.3 $\mu$m the spectra also show the CO first overtone bands for L's fade as the temperature decreases towards the T's and the dominant carbon-containing compound becomes methane (Figure 9). Water is present in all the L's and strengthens towards the T's (Figure 9). More recently spectra to longer wavelengths have become available from the Spitzer Space Telescope \cite{cushing2} and they clearly show the umbrella mode of ammonia $\nu_2$ at 933 cm$^{-1}$ (10.7 $\mu$m) in T dwarfs (Figure 10).

Unlike the situation for water \cite{bt2},
reliable molecular opacities for hot methane and hot ammonia are not available. Hot emission spectra of methane at 800, 1000 and 1273
K were recorded to match the range of temperatures encountered in T dwarfs \cite{nassar}. These
spectra are not assigned so it is difficult to use them except at
the temperatures at which they were recorded. Cushing et al. \cite{cushing} have used our hot laboratory spectra to identify features
due to methane in T dwarf spectra.

The best existing database for methane and ammonia is HITRAN 2008 \cite{hitran}, which is aimed at the calculation of infrared transmission of the Earth's atmosphere at 300 K. HITRAN is missing the hot bands needed for high temperature sources and is also rather incomplete in the near infrared. New laboratory spectra for hot methane and ammonia are clearly needed and some work is underway in our research group. There is also steady progress on the calculation of the
methane spectrum using \emph{ab initio} methods \cite{ch4calc}, but the current
predictions are far from spectroscopic accuracy; similar theoretical work has been started on ammonia \cite{nh3calc}. Hot ammonia is particularly significant for the next cooler spectral class---Y brown dwarfs---that have surface temperatures less than the T's \cite{ydwarf2}.

\section{AGB, S and C Stars}

The spectra of S- and C-type stars are complicated because of the variety of molecular species that can form as the C/O abundance ratio increases as stars evolve off the main sequence and start to burn He. S-type stars have C/O of about 1 while C-type stars have C/O greater than 1 and are defined empirically by the presence of certain molecular bands in their spectra. These stars have surface temperatures that are roughly parallel to those of M stars and are believed follow the evolutionary sequence M-MS-S-SC-C, allowing for the possibility of intermediate MS and SC stages \cite{s-star1}. Values for C/O are about 0.5 for M, 0.6 for MS, 0.8 for S and 1 for SC \cite{s-star2}. As discussed in the introduction, at the end of a star's life on the main sequence stellar evolution proceeds through the red giant, AGB, planetary nebula and white dwarf stages. The empirical M-MS-S-SC-C sequence is therefore a consequence of stellar evolution and corresponds to the red giant and AGB part of stellar demise.

S-type stars are traditionally classified by the presence of ZrO and YO bands in addition to the usual TiO bands of M-stars \cite{s-star1,s-star2,s-star3}. The heavy elements Y and Zr are beyond the Fe abundance peak and are formed by \emph{s}-process neutron reactions that are characteristic of stars that have left the main sequence. The classification notation used is Sx,y or Sx/y, in which x is the usual 0 to 9 for surface temperature and y is a number from 1 to 7 based on the relative intensity of TiO and ZrO bands (which is a proxy for the increasing C/O ratio) \cite{s-star3}. A number of visible transitions of ZrO ($\beta$ system, $\mathrm{c}^3\Pi - \mathrm{a}^3\Delta$; $\gamma$ system, $\mathrm{b}^3\Phi - \mathrm{a}^3\Delta$ and $\mathrm{B}^1\Pi - \mathrm{X}^1\Sigma^+$ transition \cite{s-star1,zro1,zro2}) and YO ($\mathrm{B}^2\Sigma^+ - \mathrm{X}^2\Sigma^+$ and $\mathrm{A}^2\Pi - \mathrm{X}^2\Sigma^+$ transitions \cite{yo1,yo2}) are seen. With much of the oxygen tied up in CO, a number of unusual sulfur-containing molecules are able to form in S-type stars including TiS \cite{tis1,tis2,tis3} and ZrS \cite{zrs1,zrs2} observed in the near infrared in, for example, R Andromedae (R And, Figure 11 \cite{j-band}).

\begin{figure}
\begin{center}
\rotatebox{90}{\resizebox{!}{13cm}
{\includegraphics{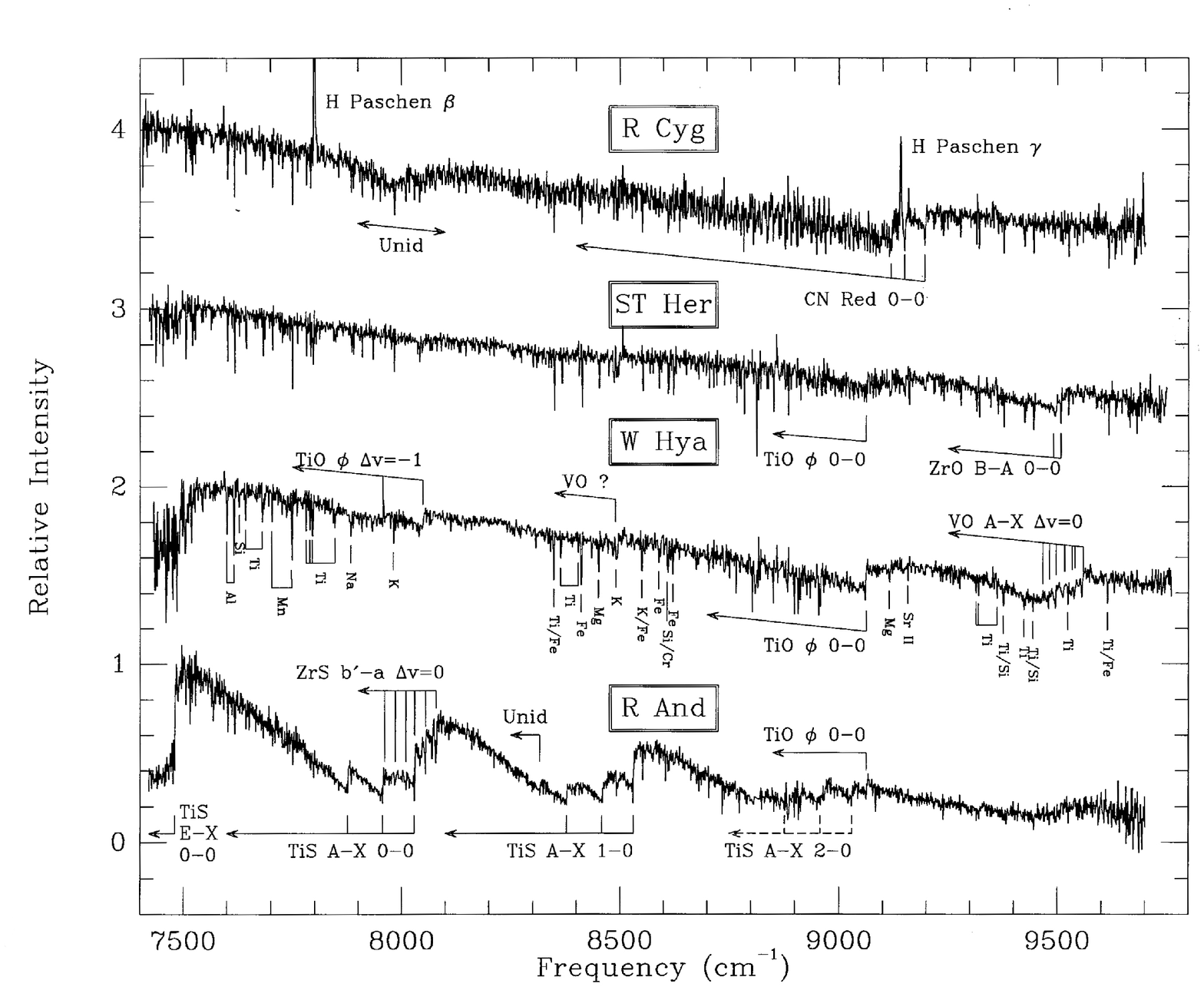}}}
\caption{Near infrared spectra of four S-type stars \cite{j-band}. Reproduced by permission of the AAS.}
\end{center}
\end{figure}

A high resolution FTS spectrum of the S-type star R Andromedae was recorded by Ridgway et al. \cite{ridgway} in the 2400--2800 cm$^{-1}$ region. Vibration-rotation lines of OH, NH, CH and HCl fundamental bands and the first overtone of CS \cite{cs1,cs2} were identified and later the fundamental band of SH \cite{sh1,sh2} was noted. To shorter wavelengths CO and the Red System of CN are very strong in $\chi$ Cygni, an S-type Mira variable that has a regular pulsing stellar atmosphere \cite{k-band}. Low resolution spectra from the Infrared Space Observatory (ISO) \cite{s-star2} suggest that HCN can be seen near 3 microns \cite{hcn1,hcn2}, and recently SiO and SiS were identified in Spitzer Space Telescope spectra near 10 microns \cite{sis}.

There are two types of S stars: intrinsic and extrinsic that can be separated by the presence of the unstable element technetium (Tc) in the photosphere \cite{jorissen}. Radioactive $^{99}$Tc (the isotope formed by neutron reactions in AGB stars) has a half-life of $2.1 \times10^5$ years and does not occur naturally on Earth. Tc, however, is an \emph{s}-process element that can be identified through its atomic lines in intrinsic S stars (i.e. genuine AGB stars). Extrinsic S stars do not have Tc because it has all decayed away. It turns out that all extrinsic S stars are binary with a white dwarf as a companion. Apparently the heavy \emph{s}-process elements such as Y and Zr in extrinsic S stars were transferred through mass loss during the companion's AGB phase.

Carbon stars have perhaps the richest of all stellar spectra and this has caused considerable confusion in their classification. Traditionally C stars are identified by the strong visible absorption of the C$_2$ Swan system, CN Red and Violet Systems and the CH A-X transition \cite{c-star1,c-star2,c-star3}. Carbon stars can also display the Merrill--Sanford bands due to the $\mathrm{\widetilde{A}^1B_2 - \widetilde{X}^1A_1}$ transition of SiC$_2$ (origin band near 497.7 nm \cite{sarre}) and the comet system of C$_3$ ($\mathrm{\widetilde{A}^1\Pi_u - \widetilde{X}^1\Sigma^+_g}$ transition near 405.2 nm \cite{comet}). The simple approach of classifying carbon stars by their surface temperature and the amount of carbon (as measured by the strength of the C$_2$ Swan bands) is not very successful because of the wide range of objects that have C/O greater than 1. The surface temperatures of C stars are notoriously difficult to determine. The current system divides C stars into three main sub-classes R, N and CH (denoted as C-R, C-N and C-H), and then further categorises each of these by temperature, strength of C$_2$, peculiar carbon isotope ratios and luminosity \cite{c-star2}. The R type is hotter and has no Ba lines from the \emph{s}-process, while the cooler N-type has enhanced heavy elements with strong blue absorption. The CH type has strong CH absorption and the J-subtype also has strongly enhanced $^{13}$C abundances, with $^{12}$C/$^{13}$C ratios as high as 2 as compared to the solar value of 92. Isotopic abundances show great deal of variation in astronomy and often provide important clues to the evolutionary history of the object. CH carbon stars are believed to be part of binary systems with white dwarfs (like extrinsic S stars) and their anomalous abundances are due to mass transfer from the other star \cite{ch-stars}.

The near infrared spectra of carbon stars are dominated \cite{loidl} by the CN Red System, the Phillips and Ballik--Ramsay ($\mathrm{b^3\Sigma^-_g - a^3\Pi_u}$ \cite{b-r-c2}) systems of C$_2$ and CO overtone bands. The vibration-rotation bands of HCN and C$_2$H$_2$ appear in the 3 $\mu$m region and to slightly longer wavelengths CH, NH and CO fundamentals and the CS first overtone can be seen in both ISO spectra \cite{aoki} and at higher resolution in FTS spectra \cite{ridgway,4mfts}. In the ISO spectra \cite{aoki} the fundamental band of CS and the first overtone of SiS are detected near 7 $\mu$m plus the $\nu_2$ bending mode of HCN at 14 $\mu$m and the $\nu_5$ bending mode of C$_2$H$_2$ at 13.5 $\mu$m \cite{aoki2}. The detection of C$_3$ has also been suggested using low resolution ISO \cite{jorg} and Spitzer spectra \cite{spitzer,lmc}.

It is also possible to have oxygen-rich AGB stars and their infrared spectra typically show OH, H$_2$O, SiO, CO, CO$_2$, SO$_2$, HCl and HF, as expected from their composition and the surrounding cloud of molecules from mass loss \cite {w-hydra,4mfts2,hf-agb}.

\section{Extrasolar Planets}

The first extrasolar planet---a `hot Jupiter' orbiting around the star 51 Pegasi---was discovered by Mayer and Queloz in 1995 \cite{esp2}, the same year the first brown dwarf was observed \cite{bd1}. The parent star is classified as G2 V and is very similar to the Sun, but the 51 Pegasi b planet is very strange indeed. (Planets are named using the parent star and lower case letters b, c, d, and so forth in order of discovery.) 51 Peg b has a mass of about 0.5 $M_{\mathrm{J}}$, but is located at 0.05 AU (1 AU is the mean Earth-Sun distance of $1.5\times10^{11}$ m) and has an orbital period of 4.2 days. In essence it would be as if Jupiter were to be transported so it was 7 times closer to the Sun than Mercury in our Solar System. The composition of 51 Peg b is expected to be similar to that of Jupiter but with a surface temperature of about 1300 K rather than about 125 K for Jupiter.

Almost all of the hundreds of extrasolar planets (or exoplanets, see http://exoplanet.eu/) have been detected by observing periodic Doppler shifts in the spectrum of the parent star. A star and a planet orbit about a common centre of mass and this motion causes the parent star to periodically move away and toward the observer. For example, the Sun-Jupiter system would have a periodic Doppler shift of 13 m/s for lines in the Sun \cite{esp}. Detection sensitivities are currently better than 3 m/s, which corresponds to a frequency shift of about 5 MHz for the Na D line. Interestingly, most of the Doppler shift observations view the star through a temperature-stabilised iodine cell (as a frequency reference) and record a high resolution visible spectrum. It is not possible to determine the inclination angle of the orbital plane from the Doppler effect, but these observations give the orbital radius, eccentricity of the orbit and a lower limit for the mass of the planet (as long as the stellar mass is known). The Doppler shifts are largest for large planets close to the parent star so there is a selection effect that favours the detection of hot Jupiters. Given a few technical improvements the Doppler shift method should be able to detect Earth-like planets.

For extrasolar planets, it is possible that the orbital inclination (the angle between the Earth-star vector and the perpendicular to the star-planet orbital plane) is close to 90$^\circ$. The planet can then transit the star and be detected by the small decrease in light as the planet passes in front of the star. The hot Jupiter HD 209458b was the first transiting planet observed with a dip in light intensity of about 1.5\% \cite{transit}. This type of precision photometry is best done from orbit and the Corot (http://smsc.cnes.fr/COROT/) and Kepler missions (http://kepler.nasa.gov/) are currently searching for Earth-like planets by their transits. The combination of Doppler and transit data allows for a rather complete description of the planet including orbit inclination, planetary mass, radius and density.

The technique of `transit spectroscopy' allows the detection of molecular absorptions in exoplanets. The idea is simple: the effective size of a planet with an atmosphere depends on the wavelength because the planet appears bigger at a strongly absorbing wavelength than at a wavelength for which the atmosphere is transparent. By measuring very precisely the size of the dips in the stellar radiation during transit as a function of wavelength, a spectrum of the planet is obtained. Swain et al. \cite{tinetti} observed the transiting exoplanet HD 189733b with the Hubble Space Telescope using the NICMOS camera in the 1.4--2.5 $\mu$m range. HD 189733b is a hot-Jupiter planet in a 2.2 day
orbit around a bright K2 V primary star. A model atmosphere of mainly molecular hydrogen with a water
relative abundance of $5\times10^{-4}$ and a methane abundance of $5\times10^{-5}$ was needed to match the observed wavelength dependence of the transit dips. They used the BT2 \cite{bt2} linelist for water and a combination of HITRAN \cite{hitran} and Nassar and Bernath \cite{nassar} for methane.

More recently the `emergent flux' from the same planet was directly detected \cite{swain} with a similar instrument configuration as used for transit spectroscopy \cite{tinetti}. The NICMOS instrument on Hubble was used again for imaging spectroscopy with a grism (a grating ruled onto a prism inserted in the field of view of an infrared camera) that gave a resolving power of 40 in the 1.4--2.5 $\mu$m range. In essence Swain et al. manipulate the spectra recorded just before the planet disappears behind the star (i.e. star plus planet) and spectra recorded during occulation with the planet behind the star (i.e. star only) to obtain the emission spectrum of the planet alone. The planetary flux contained modulations that required the addition of H$_2$O, CO and CO$_2$ to their model atmosphere. The interpretation of such spectra are not straight forward because the molecular bands could appear in absorption or emission or both, depending on the details of the atmospheric pressure-temperature height profile and the composition of the different atmospheric layers. The non-detection of methane is explained as being due to the different atmospheric regions sampled by emission spectroscopy (dayside emission from higher pressure and temperature regions) as compared to transit spectroscopy (sampling of lower pressure cooler outer regions in the terminator between night and day).

\section{Molecular Opacities}

The spectra of cool objects are rich in molecular absorption features,
which presents a major difficulty in the calculation of model
atmospheres needed to simulate the spectral energy distribution
functions. Compared to atoms, molecules have many more energy levels
and hundreds of millions of absorption lines are needed for
molecular opacities. It is not feasible to obtain millions of line
parameters from laboratory observations alone: they must be provided
by theory. Yet the purely theoretical calculation of molecular
opacities is also not satisfactory because, except for molecular
hydrogen, \emph{ab initio} quantum chemistry is not yet sufficiently
accurate. In other words, laboratory measurements alone do not
provide sufficient lines while purely theoretical calculations do
provide enough lines, but not with enough accuracy. The solution to
this problem is to extend laboratory measurements with theory to
provide the required molecular opacities. In essence, the strongest
lines seen at high resolution come from laboratory data, but the
millions of weaker lines that provide a kind of `continuum'
absorption come largely from theory. The calculation of molecular opacities has been recently reviewed by Sharp and Burrows \cite{sharp}.

Our approach to the calculation of molecular opacities for the electronic transitions of metal hydrides such as
FeH illustrates the contributions chemical physics can make in astronomy.  The simulation of these spectral energy
distributions require spectroscopic line parameters (positions,
intensities and pressure-broadening parameters) to be used in a
model atmosphere \cite{allard}.  The generation of the required
molecular opacities is not an easy task for molecules such as CrH,
FeH and TiH.  The assignment of perturbed laboratory spectra are
difficult and require considerable patience using combination
differences.  The use of just laboratory data, however, is not
satisfactory for astronomical purposes because even for metal
hydrides tens of thousands of lines contribute to the observed
stellar and substellar spectra.  The vibrational and rotational
levels generally need to be extended to higher v's and \emph{J}'s than
observed in the laboratory.  This can be done empirically or by using the predictions of \emph{ab initio} quantum chemistry. Purely
\emph{ab initio} opacities, however, are not satisfactory because the
calculated line positions are not accurate enough. A judicious
combination of experiment and theory works best to cover a wide
range of v and \emph{J}.  Minor isotopes and satellite branches should also
be included, usually from theory, because when absorption from the
main lines saturate, the weaker satellite and minor isotopologue
lines become important.

Line intensities are also required and the best approach is to
measure the radiative lifetimes for a few lines by, for example,
laser spectroscopy. The transition dipole moment function can be
computed by \emph{ab initio} quantum chemistry, but typical errors for
state-of-the-art calculations are 10-25\%. For example, our
prediction of the radiative lifetime of the A$^{6}\Sigma^+$ v=0
level of CrH was 0.75 $\mu$s \cite{crh1} and a subsequent
experimental measurement \cite{lifetime} yielded a value of
0.939 $\mu$s, some 25\% larger. Perhaps the most reliable approach
is to compute the \emph{ab initio} transition dipole moment function and
then scale it with an experimental measurement to match the
observations. Theory is thus being used to extend an experimental
value to cover a larger range of vibrational levels. Finally,
pressure-broadening parameters can also be determined by experiment
or theory, but they are difficult to measure or compute so generally
some reasonable values are just assumed \cite{crh2}.

The first step in obtaining molecular opacities for CrH, FeH and TiH
is to use the observed lines for the A$^{6}\Sigma^+$--X$^6\Sigma^+$
(CrH \cite{crh1}), F$^4\Delta$--X$^4\Delta$ (FeH \cite{phillips}),
A$^{4}\Phi$--X$^4\Phi$ (TiH \cite{tih2}) and
B$^{4}\Gamma$--X$^4\Phi$ (TiH \cite{launila}) transitions in fits
to obtain the usual spectroscopic constants for each vibrational
level. These constants are obtained by deweighting the perturbed
lines, which is particularly difficult for the heavily perturbed FeH
bands. The next step was to carry out \emph{ab initio} calculations of
molecular properties using multireference configuration interaction
(MRCI) method with the MOLPRO suite of programs \cite{crh1,feh2,tih1}.
Large atomic basis sets were used and scalar relativistic effects
were included. In the initial multireference part of the
calculation, the valence electrons were distributed among the
valence orbitals (`active space') to generate a large number of
Slater determinants that were treated equally in a CASSCF (Complete
Active Space Self Consistent Field) calculation. Additional
electronic correlation was then included by configuration
interaction (CI).  These extensive state-of-the-art calculations are
required to obtain reliable molecule properties for transition metal
hydrides.

The calculated potential energy points for each electronic state
were used to calculate vibrational wavefunctions, vibrational
energies and vibrationally-averaged rotational constants ($B_v$) by
solution of the one dimensional radial Schr\"{o}dinger equation.
This was particularly important for TiH \cite{tih1} because only the
0-0 vibrational bands are available for the A-X and B-X electronic
transitions.  For TiH, therefore, the vibrational constants were
taken from \emph{ab initio} calculations for each electronic state. The
vibration-rotation energy levels were calculated (including the
effects of spin-orbit and spin-spin coupling) for FeH \cite{feh2}
and TiH \cite{tih1} for v's from 0 to 4 or 5 and rotational levels
typically up to \emph{N} of about 50.  In our first effort on CrH
\cite{crh2}, we covered only up to v=3 and about \emph{N}=40, which
should be extended to higher v and \emph{N}. These tables of energy levels
are adjusted to contain the experimental term values derived
directly from the observations if they are available, otherwise
calculated values are used.  In this way, the experimental line
positions are reproduced by the energy levels.

In the next step, all possible transitions are computed from the
tables of energy levels, even the weak satellite lines and weak
bands not seen in the experiments. Each calculated line requires an
intensity so an Einstein $A_{ij}$ value was calculated.  The
starting point was the transition dipole moment for each electronic
transition calculated from the electronic wavefunctions.  The
vibrational wavefunctions were then used to calculate the Einstein
$A$ value for each vibrational band.  The Einstein $A$ value (in
s$^{-1}$) for each line was then calculated using the formula $A =
A_{v^{\prime}v^{\prime\prime}}HLF/(2J^{\prime}+1)$, in which $HLF$
is a H\"{o}nl--London factor (for rotation) and
$A_{v^{\prime}v^{\prime\prime}}$ is the Einstein $A$ value for the
$v^{\prime}-v^{\prime\prime}$ band. The H\"{o}nl--London factors
were obtained from the Hund's case (a) expressions (which had to be
derived for the $^4\Delta$--$^4\Delta$, $^{4}\Phi$--$^4\Phi$ and
$^{4}\Gamma$--$^4\Phi$ cases) and the rotational wavefunctions
obtained during the calculation of the energy levels. The line
strength \cite{book} of each line is most convenient in the form of
an integrated cross-section $\int\sigma d\nu$ in units of cm$^2$
s$^{-1}$ molecule$^{-1}$. The integration over frequency is just to
eliminate the line shape function.  The integrated cross-section for
a line is calculated from the Einstein $A$ by
\begin{equation}
\int\sigma d\nu=
\frac{1}{8\pi\tilde{\nu}^2Q}A(2J^{\prime}+1)\exp(-E^{\prime\prime}/kT)(1-\exp(h\nu/kT))
\end{equation}
with $\tilde{\nu}$ in cm$^{-1}$; $Q$ is the internal partition
function. As the partition function appears in the line strength
function, we also re-evaluated the thermochemistry for each molecule
using the best estimates of the dissociation energies. We computed
the partition function using estimated spectroscopic constants for
all of the low-lying spin components and electronic states. The correct evaluation of
the partition function is a frequent problem in astronomical applications and the simple rigid rotor, harmonic oscillator expressions that are approximately valid
at 300 K, are rarely reliable at high temperature.

The new
molecular opacities and thermochemistry for CrH, FeH and TiH have
been incorporated into a spectral synthesis code by Burrows
\cite{crh2,feh2,tih1}.  A comparison of the observed spectrum
of the L5 brown dwarf 2MASS-1507 and the spectral simulation
(including FeH and CrH) was carried out \cite{feh2}. The agreement
was reasonable, considering that L dwarfs are hard to model because
of the presence of dust in their atmospheres. More recent comparisons with high resolution spectra are not so favourable \cite{reiners2,tennyson} and it is clear that many of the line positions and intensities will need to be adjusted based on experimental observations.

\section{Conclusions}

Astronomy of the cool Universe is the study of molecules. As illustrated in this review, the tools of modern chemical physics can be used to great advantage in astrophysics. What is also evident is that astronomy in turn provides some unique environments for molecules and for the study of molecular processes that are rarely encountered in the laboratory. A remarkable early example was the observation of only the \emph{predissociated} lines of the $\mathrm{A}^1\Pi - \mathrm{X}^1\Sigma^+$ transition of AlH in emission in the spectrum of the MS-type variable star $\chi$ Cygni \cite{herbig}. These AlH lines are not seen in laboratory emission spectrum because of tunnelling through a potential barrier in the A state, but appear through inverse predissociation in the stellar atmosphere as Al and H atoms recombine.

\section*{Acknowledgements}

Support from the UK Engineering and Physical Sciences Research Council (EPSRC) and the NASA laboratory astrophysics program is appreciated. I am grateful for the comments on the manuscript provided by N. Allen, R. Hargreaves, J. Tandy and M. Cushing.


\label{lastpage}

\end{document}